\documentclass[journal,draftclsnofoot, onecolumn, 12pt]{IEEEtran}
\usepackage{calc}

\usepackage{multirow}
\usepackage{cite}
\usepackage{graphicx,subfigure}
\usepackage{psfrag}
\usepackage{amsmath,amssymb}
\usepackage{color}
\usepackage{soul}

\DeclareMathOperator*{\defeq}{\triangleq}

\newtheorem{theorem}{Theorem}

\newtheorem{definition}{Definition}

\newcommand{\rv}{\mathbf{r}}

\newcommand{\bit}{\begin{itemize}}
\newcommand{\eit}{\end{itemize}}

\newcommand{\bc}{\begin{center}}
\newcommand{\ec}{\end{center}}

\newcommand{\ba}{\begin{array}}
\newcommand{\ea}{\end{array}}

\newcommand{\beq}{\begin{equation}}
\newcommand{\eeq}{\end{equation}}

\newcommand{\beqn}{\begin{equation*}}
\newcommand{\eeqn}{\end{equation*}}

\newcommand{\bean}{\begin{eqnarray*}}
\newcommand{\eean}{\end{eqnarray*}}
\newcommand{\bea}{\begin{eqnarray}}
\newcommand{\eea}{\end{eqnarray}}

\def\E{\mathbb{E}}

\def\hv{\boldsymbol{h}}

\def\nv{\boldsymbol{n}}

\def\rv{\boldsymbol{r}}

\def\xv{\boldsymbol{x}}
\def\yv{\boldsymbol{y}}
\def\zv{\boldsymbol{z}}
\newcommand{\T}{{\scriptscriptstyle\mathsf{T}}}

\newtheorem{remark}{Remark}

\begin{document}
\sloppy

\title{The Capacity of Known Interference Channel}
\author{
Shengli~Zhang,~\IEEEmembership{Member~IEEE,}
Soung-Chang~Liew,~\IEEEmembership{Fellow~IEEE,}
Jinyuan~Chen,~\IEEEmembership{Member~IEEE,}
\thanks{S.~Zhang is with the Shenzhen MCSP Key Lab, College of Information Engineering,
 Shenzhen University, Shenzhen, China. S. Zhang is also with the Electrical Engineering Department, Stanford University, US. Email:zsl@szu.edu.cn.}

\thanks{S. Liew is with the Department of Information Engineering, the Chinese University of Hong Kong, Hong Kong.
Email:soung@ie.cuhk.edu.hk.}

\thanks{J.~Chen is with the Electrical Engineering Department, Stanford University, US. Email:  jinyuanc@stanford.edu.}

%\thanks{This paper was partially supported by NSFC (No. 60902016), NSF Guangdong (No. 10151806001000003), and NSF Shenzhen %(No.JC201005250034A).}
}

%%%%%%%%%%%%%%%%%%%%%%%%%%%%%%%%%%%%%%%%%%%%%

\maketitle
\thispagestyle{empty}

%%%%%%%%%%%%%%%%%%%%%%%%%%%%%%%%%%%%%%%%%%%%%
\begin{abstract}
In this paper, we investigate the capacity of known interference channel, where the receiver knows the interference data but not the channel gain of the interference data. We first derive a tight upper bound for the capacity of this known-interference channel. After that, we obtain an achievable rate of the channel with a blind known interference cancellation (BKIC) scheme in closed form. We prove that the aforementioned upper bound in the high SNR regime can be approached by our achievable rate. Moreover, the achievable rate of our BKIC scheme is much larger than that of the traditional interference cancellation scheme. In particular, the achievable rate of BKIC continues to increase with SNR in the high SNR regime (non-zero degree of freedom), while that of the traditional scheme approaches a fixed bound that does not improve with SNR (zero degree of freedom).
\end{abstract}

%%%%%%%%%%%%%%%%%%%%%%%%%%%%%%%%%%%%%%%%%%%%%%%%%%%%%%%%%%%%%%%%%%%%%%%%%%%%%%%%%%
%%%%%%%%%%%%%%%%%%%%%%%%%%%%%%%%%%%%%%%%%%%%%%%%%%%%%%%%%%%%%%%%%%%%%%%%%%%%%%%%%%
%%%%%%%%%%%%%%%%%%%%%%%%%%%%%%%%%%%%%%%%%%%%%%%%%%%%%%%%%%%%%%%%%%%%%%%%%%%%%%%%%%

\section{Introduction}
This paper investigates a general model for wireless communications with known interference. In particular, we derive an achievable rate for a {\it known-interference channel} by transforming its signal processing problem to that of a general MIMO channel.

In real wireless communications systems, a receiver is often faced with the task of decoding the data from one source amidst the interference containing data from another source. Oftentimes, the receiver has information on the data embedded in the interference, hence the term {\it known interference}.  We refer to the channel model  as the {\it known interference channel}.

The known interference channel is widely encountered in many wireless networking scenarios \cite{KIC:10}, especially heterogeneous networks\cite{Shinmin_HETNET_interference:11}. Known interference channels can generally be divided into two categories. The first category is the {\it direct known-interference channel}, where the interference data is known by direct means. For example, the interference could contain self-information. This is the case in physical-layer network coding systems \cite{Mobicom:06, ANC:07}, where the signal transmitted by a relay may contain self-information at the receivers.
 %This is also the case in a full-duplex wireless system, in which the signals received by a node contain the signal simultaneously transmitted by the node \cite{Full_duplex_MS:09}.
 This is also the case in a co-channel heterogenous network, where the signals received by a pico-cell node contain the common reference signal simultaneously transmitted by the macro cell \cite{CRS_Interference_Cancellation:12}.
 The second category is the {\it indirect known-interference channel}, where the interference data is estimated or detected by the receiver itself. An example is the interference data deduced as part of the successive interference cancellation process \cite{GBG_random_access:07}, \cite{ successive_decoding:97}.

{
In this paper, we are interested in situations in which the channel gain associated with interference channel is not known. Furthermore, to simplify design, the receiver does not estimate the channel gain either. An interesting question is to what extent optimality is traded off for simplicity in such design. As will be demonstrated in this paper, near optimal performance can be obtained.
}

Although the known-interference channel is pervasive in wireless communications systems, it has not been subjected to systematic study. To our best knowledge, the only well-known interference cancellation technique is to subtract interfering signals reconstructed with the estimated channel coefficient and the known interference information. In theoretical studies, the channel estimation is usually assumed to be perfect, with the implication that the known interference can then be totally removed \cite{ANC:07}. In reality, accurate channel estimation for mixed signals (target signal and interference) is complex and difficult \cite{TWRC_ch_esti:09}. In \cite{BKIC_JSAC:13}, we proposed a novel blind known interference cancellation (BKIC) scheme for the known-interference channel. BKIC can achieve better cancellation performance than the traditional scheme without the need to estimate the interference channel.

In this paper, we give a closed-form achievable rate of the BKIC scheme by representing it in matrix form. The reformulation has an important practical ramification: it allows us to process the signal of the known-interference channel using well-established MIMO signal processing techniques, potentially expediting the deployment of interference management technology in real wireless communications systems.

With the reformulation, we prove that the achievable rate of our scheme can approach the capacity upper bound of the known interference channel in the high SNR regime. By contrast, the achievable rate of the traditional interference cancellation scheme is much smaller.

The remainder of the paper is organized as follows. In section II, we present the system model of the known interference channel and the assumptions adopted in this paper. Section III reformulates BKIC in matrix form. Section IV derives the achievable rate of BKIC. Section V discusses the potentials of this work. Finally, Section VI concludes this paper.

\section{Known Interference Channel Model}
In this section, we present the mathematical formulation for the known interference channel. In particular, we first illustrate the known interference channel model with a relay channel setup often seen in heterogeneous wireless networks. After that, we present the abstract channel model to be investigated, followed by a definition of known interference channel capacity.

\subsection{Channel Model from a Heterogeneous Network Example }
\begin{figure}[!t]
\centering
\includegraphics[width=3.5in]{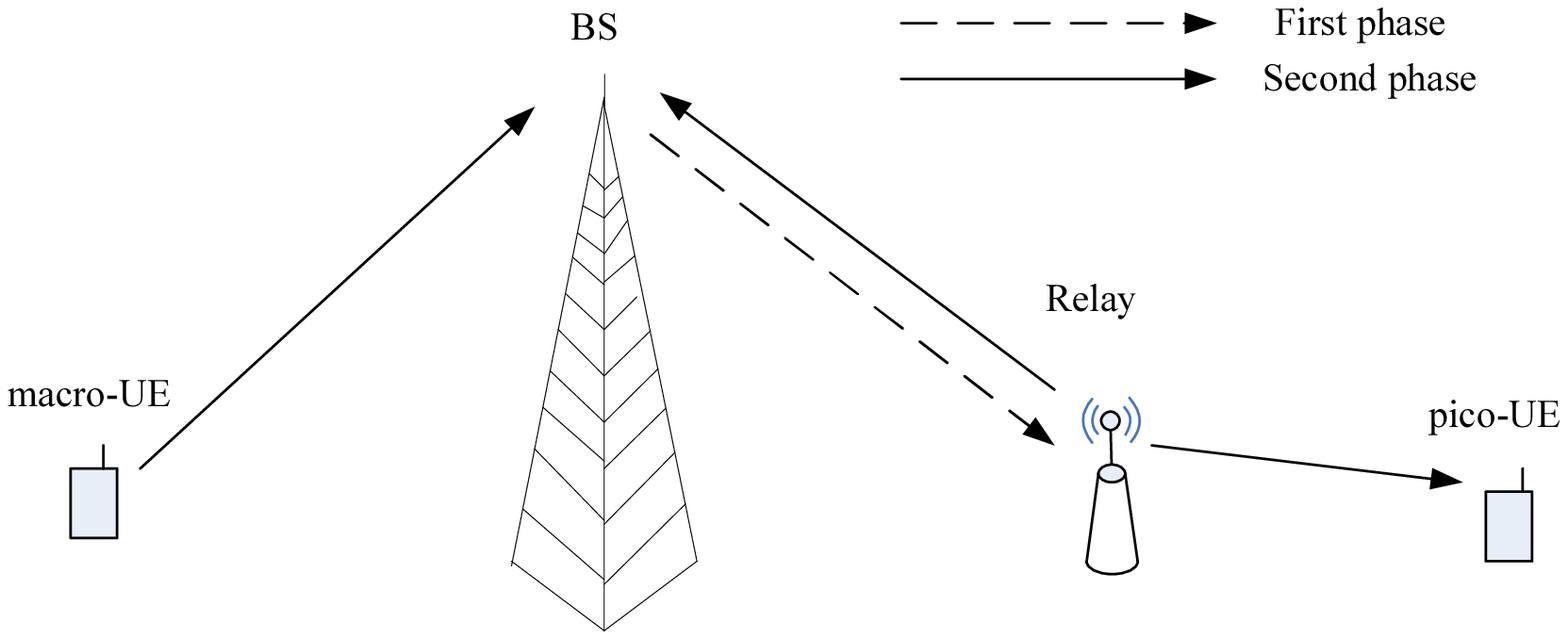}
\caption{Wireless Relay Channel}
\label{fig_relay_channel}
\end{figure}

For concreteness, we will motivate the channel model with reference to an application scenario. Consider the heterogeneous network as shown in Fig. \ref{fig_relay_channel}. Suppose that the information from the macro base station (BS) to the shown destination (pico-UE) is delivered in two phases. In the first phase, the base station transmits its data to the relay node through a backhaul channel. In the second phase, the relay forwards the data to the destination pico-UE (user equipment); at the same time, a macro-UE transmits its data to the BS through the same channel. As a result, the BS receives a superimposition of the packet from the relay and the packet from the macro-UE. The goal of the BS is to decode the data from the macro-UE amidst the known interference transmitted by the relay.

At the BS, the $k$-th received symbol in the baseband can be expressed as
\beq
\label{eq:system_model}
\begin{split}
r[k]=\sqrt{P_x}x[k]+\sqrt{P_z}h[k] z[k]+n[k]
\end{split}
\eeq
where $x[k], k = 1,2,\dots,N $  is the $k$-th signal from the macro-UE \footnote{As will be presented shortly, our BKIC algorithm does not require the knowledge of the channel gain associated with $x$. Thus, the associated channel coefficient is assumed to be 1 and it is unchanged during the whole packet length $N$}; $z[k]$  is the $k$-th known interfering signal from the relay ($z[k]$ for different $k$ are i.i.d.); $h[k]$ is the channel coefficient for the $k$-th symbol  (assumed to be constant within a block of $T$ symbols of the packet but varying between different blocks  in an independent manner with complex Gaussian distribution $CN(0,1)$ ); $P_x$ and $P_z$ are the received signal powers from the macro-UE and the interfering relay, respectively; and $n[k]$  is the complex Gaussian noise with  distribution $CN(0,\sigma^2)$ at the receiver. In general, $x[k]$ is a random variable under a total normalized power constraint as $\E  |x[k]|^2  \leq 1$. Without loss of generality, $N, T, N/T$ are assumed to be integers. %We assume $x[k]$  to have complex Gaussian distribution $CN(0,1)$, which will be explained later.
 %\footnote{ JC: I have removed the Gaussian assumption. Please define that $\E |x[k] | \leq 1$.}   %% check this comment

At the BS, the information of the interference signal $z[k]$ is known. We also assume normalized power constraint on the interference as $\E  |z[k]|^2  = 1$. Although normalized $z[k]$ is known when canceling the interference, its distribution (corresponding to the modulation scheme in real system) will also affect the cancelling performance.
%, the modulation information is needed for processing.
Our paper focuses on the known interference channel capacity with worst case distribution of $z[k]$ and our proposed BKIC scheme is independent on the distribution of $z[k]$.
%The only constraint on $z[k]$ is the total normalized power constraint as $\sum_{k=1}^N |z[k]|^2 = N$.

%Although the modulation information is known, it cannot be controlled by the receiver. Therefore, when deriving the achievable rate in Section IV, we consider the worst case modulation for $z[k]$ under the total normalized power constraint as  $\sum_{k=1}^N |z[k]|^2 = N$. As will be proved, CPM turns out to be the worst-case modulation \footnote{The worst modulation may be a little different for given known interference cancellation schemes. But CPM is no better than any other modulation for the rates discussed in this paper.}.

The SNR of the known interference channel is defined as $$\gamma=P_x / \sigma^2$$ and the power ratio between signal and interference is defined as $$\rho = P_x/P_z.$$
Throughout the paper, we use bold lowercase letters to denote vectors and the corresponding regular lowercase letters to denote elements of the vectors.

\subsection{Abstract Model and Capacity Definition}
\begin{figure}[!t]
\centering
\includegraphics[width=3.5in]{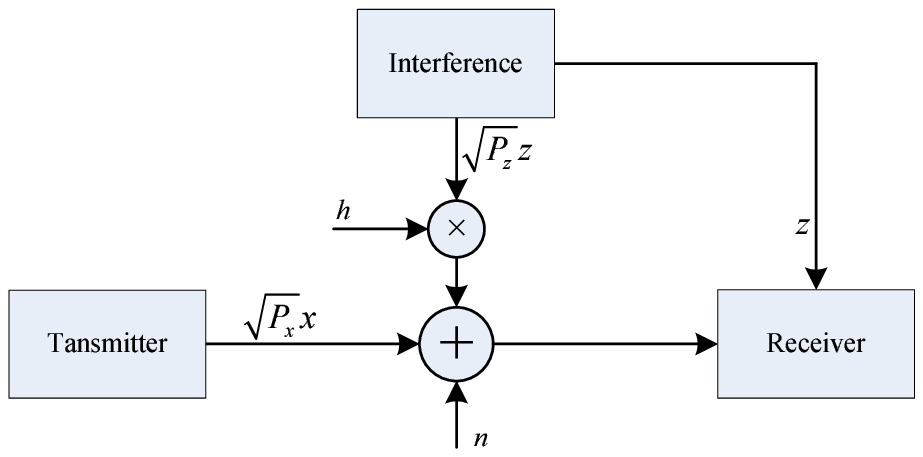}
\caption{Known Interference Channel Model}
\label{fig_KIC_channel}
\end{figure}

An abstract known-interference channel model corresponding to \eqref{eq:system_model} can be constructed as  in Fig. \ref{fig_KIC_channel}. Although simple, the model is general enough to capture various real situations, especially for our BKIC algorithm which separates the interference cancellation process and the target signal detection process. For example, even if the symbols of the target signal and interference signal were not aligned in time, the receiver could first synchronize to the interference signal during the interference cancellation process; after that, the receiver re-synchronizes to the target signal during the signal detection process . If there is carrier frequency offset between the known interference and target signal, the receiver can also synchronize to the carrier frequency of the interference first and then to the carrier frequency of the target signal next in the two successive processes. .

With reference to \eqref{eq:system_model}, a column vector form input-output relation can be written as
\beq \label{eq:system_model_vector}
\rv= \sqrt{P_x}\xv+ \sqrt{P_z}Diag (\hv) \zv+\nv
\eeq
where $\rv \defeq \left[r[k]\right]_{k=1}^{N}$, $\xv \defeq \left[x[k]\right]_{k=1}^{N}$,  $\zv \defeq \left[z[k]\right]_{k=1}^{N}$, and $\hv \defeq \left[h[k]\right]_{k=1}^{N}$ are the length-$N$ packet form of $r, x, z, n$. $Diag(\cdot)$ denotes the diagonal matrix expended with the given vector.
%$\zv^N$ can be any random sequence such that $||\zv^N||^2=\sum_{k=1}^N |z[k]|^2 =N$.

 Let $W$ to denote the transmitted message, randomly chosen from a message set with cardinality $2^{NR}$, where {$R(f_x, f_z, T, \rho, \gamma)$ } is the channel code rate. The rate $R$ is achievable if there is an encoding function to map each message to one packet $\xv$ and there is another decoding function to map each received packed, as well as the interference information, to a transmitted message as $\hat{W} $ such that the worst error probability $ P_r\left[ W\neq \hat{W} \right]$ goes to zero for very large $N$.

We now define the capacity of the known interference channel, which only depends on the interference channel coherence time $T$, the SNR $\gamma$ and the power ratio $\rho$. The capacity is independent of the input target signal distribution $f_x$ and the input interference signal distribution $f_z$. We assume that the transmitter does not have any information on the interference. It selects the distribution $f_x$ to maximize the target signal rate without considering the distribution of the interference. On the other hand, the interference is beyond the control of the transceiver, thus the worst case interference sequence, $\zv$, is of interest.

\begin{definition}{Known Interference Channel Capacity}

The capacity of known interference channel in this paper is defined as
\begin{align}
\label{eq:capacity_definition}
& C(T,\rho, \gamma)= \sup_{f_x} \left[ \inf_{f_z} R(f_x, f_z, T, \rho, \gamma) \right] \\
 & \quad s.t. \quad \E |z[k]|^2=1, \; \E  |x[k]|^2  \leq 1   \nonumber
\end{align}
\end{definition}

As will be proved later, the best distribution for $x[k]$ is Gaussian distribution. On the other hand,  the worst  distribution for interference $z[k]$ is any distribution that have constant total power for each block, which is different from the intuition that Gaussian distributed interference is worst.

\section{BKIC in Matrix Form}
With reference to \eqref{eq:system_model}, the receiver must cancel the interference part $h[k]z[k]$ so as to detect the target signal $x[k]$. Although the receiver has prior knowledge of $z[k]$, it does not know the channel coefficient $h[k]$. To cancel the interference, the conventional method is to first estimate the interference channel $h$ within a  block (recall the assumption of block fading that $h[k]=h$ is a constant within a block) and then subtract $hz[k]$  from the received signal. However, note from \eqref{eq:system_model} that the received signal also contains $x[k]$ , which is not known at this point. In particular, $x[k]$ will corrupt the estimation of $h$ even if the noise $n[k]$ is small \footnote{If the training sequence of the interference is transmitted through a channel orthogonal to the target signal, the BS can avoid this channel estimation difficulty. However, achieving such orthogonality needs coordination between the target signal transmitter and the interference transmitter, which introduces nontrivial complexity to the overall system. As shown in Appendix D, the performance of this orthogonal scheme is also limited  compared with our BKIC scheme.}.

To overcome the channel estimation difficulty, we proposed the blind known interference cancellation scheme (BKIC) in \cite{BKIC_JSAC:13} to cancel the known interference without the need for channel estimation. BKIC has near optimal performance. In this section, we reformulate it in matrix form for further insight and capacity derivation.

For simple illustration  of our BKIC scheme, in this section, we assume constant power modulation (CPM) for $z[k]$, although the scheme can be used for any possible sequence of $\zv$. Our focus on CPM here is motivated by two reasons: 1) CPM is a widely adopted modulation in real communications systems, 2) CPM
corresponds to the worst interference distribution as far as the  capacity of known interference channel is concerned as proved in the next section.

\subsection{BSIC Idea of BKIC}
Our BKIC process is divided into two steps: (i) interference cancellation; (ii) target signal recovery. In the following, we  focus on the $k$-th symbol in one data block when describing the details of BKIC processing \cite{BKIC_JSAC:13}.

\subsubsection{Step 1: Interference Cancellation}
 Let us first pre-equalize the received signal to change the known interference into an all-one sequence. Since the interference $z[k]$ is a known PSK signal with unit power (constant power modulation assumption), we can divide both sides of \eqref{eq:system_model} by $\sqrt{P_x}z[k]$ to obtain
\beq
\label{eq:system_model_simple}
\begin{split}
r'[k]&=r[k] / ({\sqrt{P_z}z[k]}) \\
&=\rho x[k]/z[k] + h[k] + n[k] / ({\sqrt{P_z}z[k]})\\
&\defeq x'[k] + h[k] + n'[k]
\end{split}
\eeq
Since there is a one-to-one mapping between $x[k]$ and $x'[k]$, our target becomes to recover $x'[k]$ hereafter in this section. By subtracting each symbol from the previous symbol, we can cancel the known interference as follows:
\beq
\label{eq:cancellation}
\begin{split}
y[k]&=r'[k]- r'[k+1] \\
&=x'[k] -x'[k+1] +  n'[k] - n'[k+1] -\Delta[k]
\end{split}
\eeq
where  $\Delta[k]=h[k+1]-h[k]$ denotes the channel variation.  With our assumption of block fading, $\Delta[k]=0$  within one data block, and it is the difference of two random Gaussian variables between two adjacent blocks %\footnote{In real system,  $\Delta[k]$ is a very small value and can be modeled by a Gaussian distribution \cite{message_passing_detection:09}, which can also be handled by BKIC \cite{BKIC_JSAC:13}.}.
Although the first step in \eqref{eq:cancellation} cancels the interference part, it also distorts the target signal and doubles the noise. We next introduce a critical step to recover the target signal without this blemish.

\subsubsection{Step 2: Target Signal Recovery}
This step aims to recover $x'[k]$ from all the post-processed $T-1$ symbols $\yv$  within one data block. We will first treat $x'[k]+n'[k]$ rather than $x'[k]$ as the target signal to be obtained. Once $x'[k]+n'[k]$  is obtain, we can then estimate $x'[k]$ using the traditional point-to-point communication method. The recovery of $x'[k]+n'[k]$ based on the observed samples in one data block, $\yv$, can be expressed as a function $f$ :
\beq
\widehat{x'[k]+n'[k]}=f(\yv)=x'[k]+n'[k]+w[k]
\eeq
where $w[k]$  is the residual interference due to incompletely removal of the known interference. In \cite{BKIC_JSAC:13}, we proposed a real valued belief propagation scheme to recover the signal optimally. For simplicity,  we use bold lowercase letters with superscripts $'$, such as $\yv',\xv',\nv'$ and $\rv'$, to denote vectors for the current block.

\subsection{Matrix Form of BKIC}
We now formulate the BKIC scheme in matrix form. Doing so provides insights that lead to more efficient signal recover algorithms. Importantly, the matrix formulation allows us to derive an achievable rate, as will be shown shortly.

In $Step \quad 1$ above,  \eqref{eq:system_model_simple} and \eqref{eq:cancellation} for all $k$ can be written in matrix form as
\beq \label{eq:cancellation_matrix}
\yv'=Q\rv' =Q(\xv'+h\mathbf{e}+\nv')=Q(\xv'+\nv')
\eeq
where $\mathbf{e}$ is a all-one length-$T$ vector and  the $(T-1)\times T$ interference cancellation  matrix Q is given by
$$
Q=\begin{bmatrix}
1 & -1 & 0 & \cdots & 0 \\
0 &1 & -1 & \cdots & 0 \\
\vdots &\ddots & \ddots & \ddots & \vdots \\

0 & 0 & \cdots & 1 & -1
\end{bmatrix}$$.

$Step \ 2$ is equivalent to recovering the vector $\xv'$  from $\yv'$.  A number of different recovery schemes that incur no information loss are possible.   An example is the real-valued belief propogation scheme in \cite{BKIC_JSAC:13}. Different from the treatment in \cite{BKIC_JSAC:13}, this paper transforms \eqref{eq:cancellation_matrix} into a standard MIMO form in order to exploit the abundant MIMO detection schemes to recover $\xv'$.

First, applying standard SVD decomposition on matrix $Q$ , $Q=USV$, we can rewrite \eqref{eq:cancellation_matrix} as
\beq
\label{eq:svd}
\yv'=Q(\xv'+\nv')=USV\xv'+USV\nv'
\eeq
where $U$ is a $(T-1)\times(T-1)$ unitary matrix,  $S$ is a $(T-1)\times T$ matrix that consists of a $(T-1)\times(T-1)$ diagonal matrix $S_1$ and an all-zero vector in the last column, and $V$ is a $T\times T$ unitary matrix. Multiplying $U^*$, the conjugate transpose of $U$, on both sides of \eqref{eq:cancellation_matrix}, we obtain
\beq
\label{eq:svd_1}
U^*\yv'=SV\xv'+SV\nv'=SV\xv'+S\nv''
\eeq
where $\nv''=V\nv'$ is a new Gaussian noise vector with the same distribution as $\nv'$. Let $V_1$ be the matrix $V$ with the last row removed. Since the last column of $S$ contains zeros only, $SV=S_1V_1$. In addition, remove the last element of $n''$ and let $\tilde{\nv'}$ be the resulting length-$(T-1)$ vector. We can then write \eqref{eq:svd_1} as
  %With $S_1$, we can also remove the last row of $V$ to obtain a new matrix $V_1$ and remove the last element of $\nv_2$ (without ambiguity, we still use $\nv_2$ %to denote such a length-$(T-1)$ noise vector). Then, we can rewrite \eqref{eq:svd_1} as
\beq
\label{eq:svd_2}
U^*\yv'=S_1V_1\xv'+S_1 \tilde{\nv'}.
\eeq
Multiplying the inverse of the full rank diagonal matrix $S_1$ on both sides of \eqref{eq:svd_2}, we can obtain
\beq
\label{eq:svd_3}
\yv''=S_1^{-1}U^*\yv'=V_1\xv'+\tilde{\nv'}.
\eeq
The above equation is equivalent to a standard MIMO channel with $T$ transmit antennas and $T-1$ receive antennas, where $V_1$ is the effective channel matrix and  $\tilde{\nv'}$ is the effective Gaussian noise. With formulation \eqref{eq:svd_3}, general MIMO detection algorithms can then be used to estimate $\xv'$ from $\yv''$. These algorithms include optimal sphere detection \cite{Turbo_code_capacity:03}, space-time trellis decoding (in traditional MIMO system) \cite{Space_time_code:98}, and suboptimal zero forcing detection. Unless otherwise stated, the term BKIC is used to refer to schemes associated with optimal recovering algorithms (i.e., algorithms that do not incur information loss in the recovery process).

In Appendix B, %\ref{proof_link}},
we present a  straightforward but suboptimal recovery scheme using the above matrix formalism. This suboptimal scheme corresponds to the traditional KIC scheme.

\section{Capacity of Known Interference Channel}
In this section, we analyze the capacity of the known-interference channel as defined in \eqref{eq:capacity_definition}. In particular, we first present a tight upper bound for the capacity. Then, we calculate the achievable rate of the proposed BKIC scheme, which can approach the upper bound. For comparison purpose, we also present the achievable rate of the traditional cancellation scheme.  It is worthwhile to restate some important assumption that Note that the interference $z[k]$ is assumed to be i.i.d for different $k$ and the channel coefficient $h[k]$ is Gaussian distributed. %To make the discussed rate (including the achievable rate and the rate upper bound) in this section independent of the modulation schemes, the transmitter selects the modulation to maximize the point-to-point channel rate since the existence of the interference could be unknown to the transmitter. Therefore, the Gaussian modulation is admitted by the transmitter. On the other hand, the interference is beyond the control of the transceiver, the worst case interference distribution, is derived and assumed in this section, under the condition of Gaussian distributed target signal. Since this section focuses on the capacity of Gaussian channel, the receiver known channel coefficient $h_x$ is assumed to be unit. Hence, the achievable rate of original objective signal $x_a$ is the same as that of $x$ .

\subsection{Upper Bounds}
This part provides a tight upper bound for the capacity of the known-interference channel. Before that, let us first review a straightforward upper bound. In the known-interference channel, if the channel coefficient of the interference  $h[k]$ is perfectly known, the interference can be exactly reconstructed and completely removed by simple subtraction. Then, the remaining signal is a traditional point-to-point channel without any interference.
%, whose capacity serves as an upper bound for that of the known-interference channel.
Thus, a straightforward upper bound is given by
\beq
\log(1+SNR)=\log(1+P_x/\sigma^2)=log(1+\gamma)
\eeq

As will be shown, this upper bound is not tight, especially in the high SNR regime. We now present a tighter upper bound of the known interference channel capacity.
\begin{theorem}
\label{th:upper_bound}
The capacity of the known-interference channel is upper-bounded by
\beq \label{eq:tighter_upperbound}
\begin{split}
C_u&=(1-\frac{1}{T})\log(1+\frac{P_x}{\sigma^2})+\frac{1}{T} \log(1+\frac{P_x}{\sigma^2+TP_z})\\
&=(1-\frac{1}{T})\log(1+\gamma)+\frac{1}{T} \log(1+\frac{\rho \gamma}{\rho+T \gamma})
\end{split}
\eeq
\end{theorem}
The detailed proof of this upper bound can be found in Appendix A. In the proof, we first argued that Gaussian distributed $x$ maximized the target rate. On the other hand, the rate is minimized by interference with fixed total interference power in each block..

The upper bound in \eqref{eq:tighter_upperbound} indicates that the block length $T$ affects the known interference cancellation as a pre-log factor $1-1/T$. This point can be understood as follows. When  $T=1$, the upper bound of the capacity is $\log(1+\frac{P_x}{\sigma^2+P_z})$ since the unknown channel coefficient makes the interference equivalent to Gaussian noise. When $T$ increases, the additional $T-1$ received signals does not induce any new unknown channel coefficients \footnote{the term $hz[k]$ can be canceled as long as the rate of $x[1]$ is small so that it can be correctly decoded and removed to obtain the good estimate of $h$. }, and a reasonable upper bound of the additional capacity is $(T-1)\log(1+\frac{P_x}{\sigma^2})$. Therefore, it is logical that the coherence time $T$ affect the upper bound as a pre-log factor $1-1/T$ for one channel use.

For a further understanding, let us write out the mutual information in one block,  $I(\xv';\rv' | \zv')$, to serve as the upper bound. Expending $I(\xv';\rv' |zv')$ as $I(\xv',h;\rv' | \zv') - I(h;\rv'|\zv'\xv')$=$I(\xv',h;\rv' | \zv') - I(h;\rv'-\xv'|\zv')$,  the first term should have a pre-log coefficient $T$ because $\xv', \rv'$ are both length-$T$ vectors while the second term should have a  pre-log coefficient $1$ since $h$ is a scalar variable. Specifically,  by evaluating the  two mutual information formulas, we can obtain the exact upper bound as $T\log(1+P_x/\sigma^2)+\log(1+TP_z/(\sigma^2+P_x))-\log(1+TP_z/\sigma^2)$, where the first two terms corresponding to the mutual information $I(\xv,h;\rv,\zv)$, and the third term corresponding to the mutual information $I(h;\rv,\zv|\xv)$.
%The detailed proof can be found in Appendix A.{\color{red} \ref{proof_upperbound}}.

%This upper bound can be understood by considering one block. The information about  $\xv$ with observation of $\rv$ and $\zv$, $I(\xv;\rv,\zv)$, is the difference %of two mutual information formulas: $I(\xv,h;\rv,\zv) - I(h;\rv,\zv|\xv)$,

As given in \eqref{eq:tighter_upperbound}, the first term of the upper bound is independent of the interference power  $P_z$; only the second term, which is much smaller compared to the first part (in the high SNR regime or with long block length $T$), depends on  $P_z$. Specifically, when  $P_z$  or $T$ tends to infinity, the second term approaches  zero. This upper bound (very tight as shown later) means that although larger known interference power will degrade the capacity of the known interference channel,  its effect is very limited.

When all the power of the interference concentrates on one symbol in each block, the upper bound in \eqref{eq:tighter_upperbound} can be achieved directly. In general case with finite $T$ and $\gamma$,  the upper bound in \eqref{eq:tighter_upperbound} cannot be achieved exactly even by the best known scheme BKIC.

\subsection{Achievable Rate with Traditional KIC}
Before treating BKIC, let us first present the achievable rate using the traditional KIC scheme. For traditional KIC with least-square channel estimation, the residual interference due to channel estimation error is treated as pure noise and the achievable rate is as follows:

\beq \label{eq:traditional_rate}
\begin{split}
R_t&=\log(1+SINR)=\log(1+\frac{(T-1)P_x}{P_x+T\sigma^2})\\
&=\log(1+\frac{(T-1)\gamma}{T+\gamma})
\end{split}
\eeq

With reference to \eqref{eq:app_sinr} in Appendix C,%\ref{proof_tradtional},
we can easily obtain the SINR of the traditional KIC scheme with CPM interference as the worst case. The detailed derivation can be found in Appendix C %\ref{proof_tradtional}
\footnote{In the Appendix D, %\ref{proof_orthogonal},
we present another popular cancellation scheme with coordinated orthogonal training sequence. Even with such cost of coordination, the achieve rate is still strictly less than the rate of BKIC.}.

In \eqref{eq:traditional_rate}, the signal power $P_x$ appears not only in the numerator but also in the denominator of $\frac{(T-1)P_x}{P_x+T\sigma^2}$  because the target signal is regarded as Gaussian noise when estimating the interference channel coefficient.

\begin{remark}
The achievable rate of the traditional KIC approaches $\log(T)$  as  the SNR $\gamma$ goes to infinity. Note that $\log(T)$ is a constant independent of the signal SNR.  More specifically, the degree of freedom of the known-interference channel using the traditional KIC processing is zero. In other words, as the SNR goes to infinity, there is a huge gap between the achievable rate of the traditional KIC and the upper bound.
\end{remark}

\begin{remark}
The achievable rate of the traditional scheme in \eqref{eq:traditional_rate} approaches the upper bound $C_u$, when the SNR is fixed and the block length $T$ goes to infinity. This means that the traditional KIC scheme is near optimal with very large block length (coherent time).
\end{remark}

\subsection{Achievable Rate with BKIC}
We now derive the achievable rate of the proposed BKIC scheme. With optimal signal recovery in BKIC, we have the following closed form achievable rate.

\begin{theorem} \label{main_theorem}
For BKIC, the achievable rate is $$(1-1/T)\log(1+\gamma)$$, which is achieved with Gaussian distributed $x$ and is independent of the interference distribution.
\end{theorem}

\begin{IEEEproof}
We first prove this theorem for the case that $z[k]\neq 0, \forall k $.

%Ignoring the pre-processing in \eqref{eq:system_model_simple}, eq. \eqref{eq:cancellation_matrix} becomes $\yv'=Q\rv=Q(\xv+h\ev+\nv)$ , where the cancellation matrix $Q$ becomes a general  matrix related to $\mathbf{z}$ , which is
In (4), if we did not perform the divide-by-z[k] pre-processing, then the corresponding matrix Q in (7) would be
$$
Q=\begin{bmatrix}
z[2] & -z[1] & 0 & \cdots & 0 \\
0 &z[3] & -z[2] & \cdots & 0 \\
\vdots &\ddots & \ddots & \ddots & \vdots \\

0 & 0 & \cdots & z[T] & -z[T-1]
\end{bmatrix}$$
It is easy to verify that matrix $Q$ has full row rank.

Then, the processing from \eqref{eq:svd} to \eqref{eq:svd_3} can also be applied to obtain the standard MIMO form as in \eqref{eq:svd_3}. It is easy to verify that the noise term $\tilde{\nv'}$ is independent of the signal $\xv$. Moreover, the effective channel state information (CSI) $V_1$ is not known to the transmitter (UE in Fig. 1) even if it knew the real channel $h_x$ , because $V_1$ depends on the interference information  $\mathbf{z}$. The capacity of the MIMO channel without channel state information at the transmitter side as in \eqref{eq:svd_3} is well established  \cite{TEL:99, Turbo_code_capacity:03}. It is
\beq \label{eq:MIMO_rate}
\begin{split}
C_{MIMO}&=\log\left( det(I_{T-1}+\frac{P_x}{\sigma^2}V_1V_1^*)\right) \\
&=(T-1)\log(1+\frac{P_x}{\sigma^2})
\end{split}
\eeq
where $I_m$ denotes the $m\times m$ identity matrix, the signal $x$ has Gaussian distribution and it is independent of the interference distribution. The second equality in \eqref{eq:MIMO_rate} is obtained by noting that the rows of $V_1$ are orthogonal with each other. As the packet length $N$ goes to infinity, the MIMO capacity in \eqref{eq:MIMO_rate} is achievable. Therefore, the achievable rate per symbol with BKIC is
\beq \label{eq:bkic_rate1}
\begin{split}
R_{BKIC} = C_{MIMO}/T=(1-1/T)\log(1+\gamma)
\end{split}
\eeq

We now prove this theorem for the case that some interference symbols $z[k]$ have zero value. Without loss of generality, we assume the last $m$ interference symbols have zero power, i.e., $z[k]=0, k=T-m+1, T-m+2,\dots, T$ . Then, the receiver only performs BKIC cancellation scheme for the first $T-m$ symbols, and we can obtain a sum rate of $(T-m-1)\log(1+\gamma)$ as in \eqref{eq:bkic_rate1}. For the last $m$ symbols, there is no interference and we can obtain a sum rate of $mlog(1+\gamma)$.  As a result, the per symbol mutual information is also
$$
R_{BKIC} = (T-1)\log(1+\gamma)/T=(1-1/T)\log(1+\gamma).
$$
This completes the proof of Theorem \ref{main_theorem}.
\end{IEEEproof}

In \eqref{eq:bkic_rate1}, the $1/T$ pre-log loss is due to the unknown of the interference channel, which is similar as the unknown channel penalty as in \cite{zheng_noncoherent_multiple_antenna_channel:02}. More specifically, recall the BKIC processing in a block. There is some information loss in Step 1 by transforming the $T$ received symbol into $T-1$ symbols. Since there is no information loss in Step 2, we can expect to obtain  $(T-1)\log(1+\gamma)$ information finally for one block. By comparing the two achievable rates in \eqref{eq:traditional_rate} and \eqref{eq:bkic_rate1}, we have the following corollary.

$Corollary:$ For the known interference channel, the achievable rate with BKIC scheme is always larger than that with the traditional KIC scheme as long as $T>1$.

\begin{IEEEproof}
Both $R_t$ and $R_{BKIC}$ are increasing functions of the SNR $\gamma$. However, their increasing rates are different, and the difference is
\beq
\begin{split}
\alpha &= \frac{\partial }{\partial \gamma}(R_t) - \frac{\partial }{\partial \gamma}(R_{BKIC})\\
&=-\frac{(T-1)\gamma}{T(1+\gamma)(T+\gamma)} < 0
\end{split}
\eeq
which means that $R_{BKIC}$ increases faster than $R_t$ when the SNR  $\gamma$ increases. Therefore,  $R_t/R_{BKIC}$ achieves the largest value as $\gamma$  goes to zero. According to the L'Hopital's rule, we have
\beq
\begin{split}
\lim_{\gamma \to 0} R_t/R_{BKIC}=1
\end{split}
\eeq
which implies that $R_t/R_{BKIC}$  is always less than 1 for non-zero SNR. Therefore, we can conclude that  $R_{BKIC}$ is always larger than $R_t$. Moreover, the gap between $R_{BKIC}$ and the traditional rate $R_t$ goes to infinity as SNR increases, because
\beq
%\[ \int_0^{\infty} \alpha \,d\gamma. \]
\begin{split}
    \int_{0}^{\infty} \alpha \, d\gamma &=\int_{0}^{\infty} \frac{\partial }{\partial \gamma}(R_t-R_{BKIC})\, d\gamma \\
     &= R_t(\gamma=\infty)-R_{BKIC}(\gamma=\infty)=-\infty
\end{split}
\eeq
This completes the proof.
\end{IEEEproof}

\begin{remark}
As SNR goes to infinity, the achievable rate of BKIC  also goes to infinity, for any given value of $T$. On the other hand, the achievable rate of traditional KIC becomes a constant as the SNR goes to infinity.
\end{remark}

\subsection{Capacity in High SNR Regime}
We now discuss the relation between the achievable rate $R_{BKIC}$ and the upper bound $C_u$ in the high SNR regime ($P_x$  increases to infinity and $\sigma^2$  keeps constant). The gap between $C_u$ and  $R_{BKIC}$ can be expressed as
\beq \label{eq:rate_gap}
\begin{split}
C_u-R_{BKIC} &= \frac{1}{T}\log(1+\frac{\rho \gamma}{\rho+T\gamma})\\
&=\frac{1}{T}\log(1+\frac{P_x}{\sigma^2+T P_z})
\end{split}
\eeq

We first discuss the high interference regime where $P_z$ goes to infinity.
When $P_z$  increases faster than  $P_x$ ( $\rho$ goes to zero), the gap in \eqref{eq:rate_gap} goes to zero.  In other words, $R_{BKIC}$  approaches the upper bound $C_u$ , which is then the capacity of the known interference channel in this case. The intuition is that when $P_z$ is much larger than $P_x$, one single symbol (e.g., the first symbol) contains almost no information of $x$ and this symbol can only help other symbols to cancel interference. Therefore, only $T-1$ effective channel uses as $R_{BKIC}$.
When $P_z$ and  $P_x$ are of the same order  ( $\rho$ keeps constant) or $P_z$  increases more slowly than $P_x$ ( $\rho$ goes to infinity), the gap is upper bounded by the constant $1/T\log(1+\rho/T)$ by ignoring $\sigma^2$ in  \eqref{eq:rate_gap}. As SNR goes to the infinity, $R_{BKIC}$  also goes to infinity and this constant gap is negligible.

We now discuss the the low interference regime where $P_z$ does not go to infinity.  When $P_z$ is constant or $P_z$ goes to 0, the gap in \eqref{eq:rate_gap} is $\log(1+\gamma)/T$, which also goes to infinity with the increase of $P_x$. In fact, when $P_z$ goes to 0 and the noise dominates the system performance, we can simply ignore the known interference in this case and achieve a rate appoaching the upper bound $\log(1+\gamma)$. In other words, BKIC is suboptimal when the known interference power is very small.

Therefore, we obtain the following conclusion:
\newline
\begin{theorem}
In the high SNR  and high interference regime, when  $\rho$ goes to zero,  $R_{BKIC}$  approaches the upper bound  $C_u$ of known interference channel with a vanishing gap; when  $\rho$ is a constant or $\rho$ goes to infinity,  $R_{BKIC}/C_u$ approaches 1. In the high SNR and low interference regime, when $P_z$ is constant or goes to 0,  $R_{BKIC}/C_u$  is more than $1-1/T$.
\end{theorem}

\begin{figure}[!t]
\centering
\includegraphics[width=3.5in]{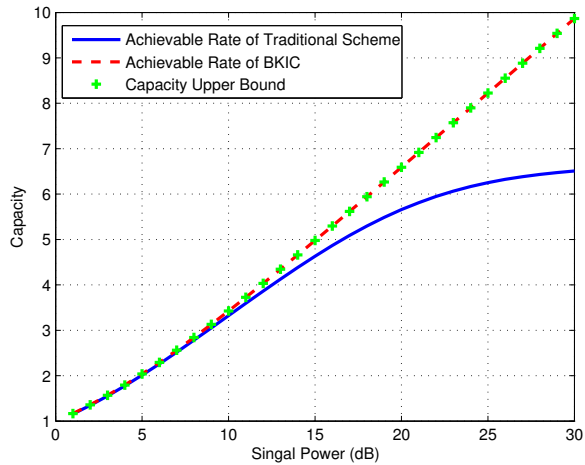}
\caption{Rate comparison with different signal power, where $P_z=P_x$, $T=100$ and $\sigma=1$ }
\label{fig_simu_power}
\end{figure}

\begin{figure}[!t]
\centering
\includegraphics[width=3.5in]{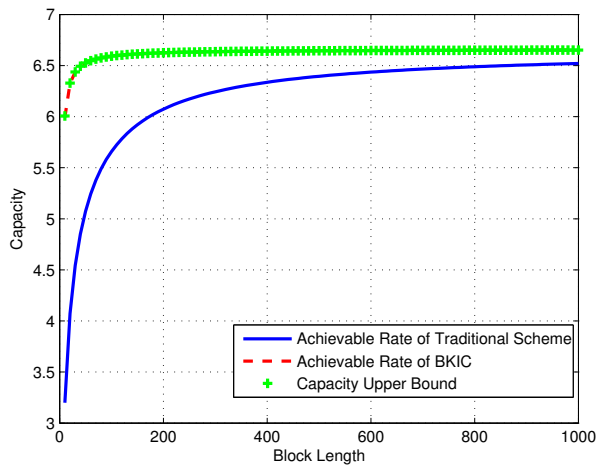}
\caption{Rate comparison with different block length, where $P_z=P_x=20dB$ and $\sigma=1$}
\label{fig_simu_length}
\end{figure}

For an intuitive comparison, we plot the achievable rates $R_t$, $R_{BKIC}$ and the upper bound $C_u$ in Fig. \ref{fig_simu_power} and Fig. \ref{fig_simu_length}, where the signal power $P_x$ is set to equal the interference power $P_z$ and the one dimensional noise variance $\sigma$ is set to 1. In Fig. \ref{fig_simu_power}, $P_x$ changes from 1dB to 30dB and the block length $T$ is fixed to 100. In Fig. \ref{fig_simu_length}, $P_x$ is fixed to 20dB and $T$ changes from 10 to 1000. We can see that our achievable rate of $R_{BKIC}$ almost overlaps with the tighter upper bound $C_u$, while the achievable rate of $R_t$ is much lower.


\section{Discussion}
This section discusses some further research issues of interest.

\subsection{Achievable Rate for General Channels}
\subsubsection{Continuous Fading Channel}
In general, the channel coefficient changes with time in a continuous way. For this type of continuous fading, the $\Delta[k]$  in \eqref{eq:cancellation} is non-zero and its distribution can be modeled as Gaussian distribution with zero mean and given variance $\sigma_{\Delta}^2$  as in \cite{message_passing_detection:09}. In this case, eq. \eqref{eq:svd} can be written as
\beq
\label{eq:svd_continue}
\yv'=Q(\xv'+\nv')+\mathbf{\Delta}=USV\xv'+USV\nv'+\mathbf{\Delta}
\eeq
Then, the processing from \eqref{eq:svd_1} to \eqref{eq:svd_3} can be applied and we obtain
\beq
\label{eq:svd_3_continue}
\yv''=S_1^{-1}U^*\yv'=V_1\xv'+\nv''+S_1^{-1}\mathbf{\Delta}.
\eeq

In block fading channel, the achievable rate of BKIC is independent of the modulation of the interference signal. In \eqref{eq:svd_3_continue}, the inverse of $S_1$ , determined by the modulation of the interference signal, will affect the distribution of the "noise term" $S_1^{-1}\mathbf{\Delta}$ , hence the achievable rate. When constant power modulation is used by the interference, the matrix $S_1$  is fixed with the pre-processing in \eqref{eq:system_model_simple}. Then the achievable rate can be easily calculated. For general modulations, derivation of the achievable rate is challenging.

\subsubsection{MIMO Channels}
Multiple input and multiple output (MIMO) is a  key technology to increase the spectrum efficiency in current wireless systems. However, the multiple channels in MIMO brings new challenges in channel estimation, especially for the multiple known interference channel as in our paper. Therefore, extending BKIC to incorporate MIMO channel is of great interest.

\subsection{Advanced BKIC Algorithms}
For BKIC, the critical step is to recover the target signal $\xv$ from the vector $\yv'$  as in \eqref{eq:cancellation_matrix}. With reference to \eqref{eq:svd_3}, the post-processed signal, $\yv''$ ,  can be regarded as the output signal of a standard MIMO channel, and the clumsy MIMO detection schemes can be applied. Some schemes thereof are discussed as follows.

An artificial suboptimal BKIC-ZF detection scheme is  discussed in Appendix B, %\ref{proof_link},
which is equivalent to the traditional KIC scheme. Moreover,  if one symbol of $\xv$ in \eqref{eq:svd_3} is pre-known by the receiver, it then is a common $(T-1)\times (T-1)$  MIMO system and the simple MIMO ZF detection scheme can be used directly. With more powerful MMSE detection rather than ZF detection, better performance can be expected. Other practical MIMO detection methods, such as BLAST MIMO, lattice reduction can also be applied, which need more discussion.

Equations \eqref{eq:svd_3} are underdetermined, although we can obtain the most probable solution for them with the signal recover schemes as in BKIC. When considering the redundancy in channel code, the original information is not underdetermined any more. Therefore, it would be interesting to design the channel decoding and signal recovery algorithms in a joint way.

\subsection{Applications of BKIC}
As discussed in the introduction part, BKIC scheme needs the interference information, which can be obtained by the receiver through direct or indirect means in many scenarios. In fact, BKIC only requires the relative amplitude information between adjacent interference symbols, not the exact information of every interference symbol. Therefore, BKIC can also be used in cases without exact interference information. For example, if $z[k]$ is spread by a known spreading sequence as in a spectrum spreading system, we can apply BKIC to cancel $z[k]$ symbol by symbol at the chip level (spreading sequence level) without knowing the value of $z[k]$.

On the other hand, with the near upper-bound achievable rate of BKIC in \eqref{eq:bkic_rate1}, we can more accurately evaluate the performance of new MAC or routing protocols specially designed to exploit the known interference cancellation at the physical layer for performance gains  \cite{overlap_trans:06, Finite_capacity_coop:09, GBG_random_access:07}.

%==================================================
\section{Conclusions} \label{sec:conclu}
We have derived a tight upper bound for capacity of a canonical known-interference channel model. The model captures many scenarios of interest in practical settings. In addition, we provide a blind-known interference cancellation (BKIC) scheme that can approach the capacity upper bound in the interference-limited regime when SNR is high. The BKIC scheme is amenable to simple implementation and we believe it can be easily incorporated into many practical communications systems.

\section *{Appendix}
\subsection{Proof of Theorem 1} \label{proof_upperbound}
For the model expressed in \eqref{eq:system_model_vector}, the receiver observes the received signal $\rv$  and the interference data $\zv$, and tries to detect the target information $\xv$, with unknown interference channel coefficients. We first derive the upper bound of the achievable rate with any given distribution of interference. After that, the worst case of interference and the corresponding upper bound are obtained.

Recall that $W$ denotes the transmitted message, and $\rv$, $\zv$, and $\hv$ denote the received signals, interference signals and the channel gains over  $N$ consecutive channel uses.

For notational convenience, we rewrite the channel model   \eqref{eq:system_model_vector}  in the  form of a block model\footnote{$N, T, N/T$ are assumed to be integers, and the time index $t=1$ is the first symbol period of the first block.} with $i$ as the block index
 \[\rv_i  = \sqrt{P_x} \xv_i + \sqrt{P_z} h_i \zv_i  +  \nv_i,  i=1,2,\cdots, N/T\]  where $\rv_i \defeq \bigl[ r[ iT-T +1 ], r[ iT-T +2 ], \cdots, r[iT]  \bigr]^\T$, $\xv_i \defeq \bigl[ x[ iT-T +1 ], x[ iT-T +2 ], \cdots, x[iT]  \bigr]^\T$, $\zv_i \defeq \bigl[ z[ iT-T +1 ], z[ iT-T +2 ], \cdots, z[iT]  \bigr]^\T$, , and $h_i$ is a scalar denoting the channel coefficient for block $i$.   In what follows, we assume $N$ and $N/T$ to be sufficiently large.
Starting from Fano's inequality, we have
\beq \label{eq:fano_inequality}
\begin{split}
 NR &= h(W)\\
    &= h(W) -  h (W | \rv, \zv)  +  h(W | \rv, \zv )\\
    &= I(W; \rv, \zv)  +  h(W | \rv, \zv)\\
    &\leq   I (W; \rv, \zv)  + N \epsilon_N
\end{split}
\eeq
where R denotes the achievable rate and the first equality is from the definition of the entropy $NR = h ( W )$ , the third equality follows from the mutual information definition  $h (W) -  h (W | \rv, \zv)= I (W; \rv, \zv) $ ,  the last step follows from Fano's inequlity, i.e.,  $h(W | \rv, \zv) \leq  N \epsilon_N$  (the error detection parameter $\epsilon_N$  goes to zero when $N$ goes to infinity). Then, we can rewrite the upper bound in \eqref{eq:fano_inequality} as
\begin{align}
&NR -N \epsilon_N     \nonumber\\
&\leq I(W;  \rv,\zv )     \nonumber\\
  & =  I(W; \zv ) + I(W;  \rv | \zv )       \\
 & =   I(W;  \rv | \zv )     \label{eq:3523} \\
 & =   h(\rv | \zv ) -h(\rv |W, \zv )    \label{eq:325}
\end{align}
where  \eqref{eq:3523} follows from the independence between $W$ and $\zv$.

We proceed to bound the first term in  RHS of \eqref{eq:325}:
\begin{align}
 h(\rv | \zv )    & = \sum_{i=1}^{N/T}  h( \rv_i | \rv_{1}, \cdots, \rv_{i-1}, \zv )  \label{eq:664} \\
& \leq \sum_{i=1}^{N/T}  h( \rv_i | \zv_i )  \label{eq:837}
\end{align}
where \eqref{eq:664} follows from the basic chain rule, \eqref{eq:837} uses the fact that conditioning reduces entropy.
For each term in the summation in \eqref{eq:837}, we have
\begin{align}
& h( \rv_i | \zv_i )     \nonumber\\
& =   \E_{ \zv_i }    h( \rv_i | z_i = \zv_i )    \label{eq:48342} \\
& =   \max_{ p_{\xv_i}: \  \text{trace} [\Phi] \leq  T  }   \E_{ \zv_i }    h( \rv_i | z_i = \zv_i )    \label{eq:33242} \\
& = \!\!\!  \max_{ \Phi:  \ \text{trace} [\Phi] \leq  T  }   \E_{ \zv_i }    \! \log (\pi e)^{T} \det \bigl[ P_x \Phi  + P_z \zv_i\zv_i^*  +   \sigma^2 I  \bigr]   \label{eq:9357} \\
& \leq    \E_{ \zv_i }   \log (\pi e)^{T} \det \bigl[ P_x I  + P_z\zv_i\zv_i^*  +   \sigma^2 I  \bigr]   \label{eq:2752} \\
& =   \! \E_{ \zv_i }  \!\! \log\Bigl( (\pi e)^{T} (P_x + \sigma^2)^{T} (1 +  P_z/(P_x + \sigma^2)  ||\zv_i||^2 )  \Bigr) \label{eq:24245} \\
& =     T \log (\pi e)  +  (T-1)  \log(P_x + \sigma^2) \nonumber\\  &\quad +   \E_{ \zv_i }  \log(P_x + \sigma^2 +  P_z  ||\zv_i||^2 )   \label{eq:2843}
\end{align}
where $\Phi \defeq \E [\xv_i \xv_i^{*}] $,  and \eqref{eq:9357} is from the fact that  Gaussian input is a entropy maximizer and the corresponding covariance of $\rv_i $ becomes $  P_x \Phi  + P_z \zv_i\zv_i^*  +   \sigma^2 I   $ for a given $\zv_i$,
\eqref{eq:2752} follows from the fact that equal power allocation is optimal provided that the input is independent of $\zv$, \eqref{eq:24245} follows from the Sylvester's determinant theorem that $\det(I_m+AB)=\det(I_n+BA)$.  Thus, combining \eqref{eq:837} and \eqref{eq:2843}  gives
\begin{align}
 h(\rv | \zv )    & \leq   N \log (\pi e)  + N (T-1)/T\log(P_x + \sigma^2) \nonumber\\  &\quad +  \sum_{i=1}^{N/T}  \E_{ \zv_i }  \log(P_x + \sigma^2 +  P_z ||\zv_i||^2   ) .    \label{eq:3167}
\end{align}

We now consider the second term in the RHS of \eqref{eq:325}, and have
\begin{align}
&h(\rv |W, \zv )  \nonumber\\
&= \sum_{i=1}^{N/T}  h(\rv_i |\rv_1,\cdots,\rv_{i-1}, W, \zv )   \label{eq:9472}  \\
&= \sum_{i=1}^{N/T}  h( \sqrt{P_z} h_i \zv_i  +  \nv_i   |\rv_1,\cdots,\rv_{i-1}, W, \xv, \zv )   \label{eq:5922}   \\
&= \sum_{i=1}^{N/T}  h( \sqrt{P_z} h_i \zv_i  +  \nv_i   |\zv_{i})   \label{eq:9827}   \\
&= \sum_{i=1}^{N/T}  \E_{\zv_i }     \log (\pi e)^{T} \det \bigl[  P_z \zv_i\zv_i^*  +   \sigma^2 I  \bigr]    \label{eq:76262}   \\
&= \sum_{i=1}^{N/T} \E_{\zv_i }     \log \Bigl( (\pi e)^{T} (\sigma^2  )^{T-1}  (P_z ||\zv_i||^2  +   \sigma^2  ) \Bigr)  \label{eq:48424}   \\
&=N \log (\pi e)   +N(T-1)/T  \log  \sigma^2    \nonumber\\ &\quad + \sum_{i=1}^{N/T}   \E_{\zv_i }     \log  (P_z ||\zv_i||^2  +   \sigma^2  )     \label{eq:27528}
\end{align}
where \eqref{eq:9472}  results from the basic chain rule,  \eqref{eq:5922} uses the fact that $\xv$ is a function of the message $W$,  \eqref{eq:9827} follows from the fact that $\{\rv_1,\cdots,\rv_{i-1}, W, \xv, \zv \} \to \zv_{i} \to \sqrt{P_z} h_i \zv_i  +  \nv_i  $  forms a Markov chain,  \eqref{eq:76262} stems from the fact that, given $\zv_i$,   $\sqrt{P_z} h_i \zv_i  +  \nv_i$ is a Gaussian vector with covariance being $P_z \zv_i\zv_i^*  +   \sigma^2 I$.

Finally, combining \eqref{eq:325}, \eqref{eq:3167} and \eqref{eq:27528} gives
\begin{align}
&R -\epsilon_N     \nonumber\\
&\leq \!\! \frac{  (T-1)}{T} \log(P_x + \sigma^2) \!+\!  \frac{1}{N} \sum_{i=1}^{N/T} \E_{ \zv_i }  \log(P_x + \sigma^2 +  P_z ||\zv_i||^2   )   \nonumber\\  &  \quad -\frac{ (T-1)}{T}  \log  (\sigma^2  )  -  \frac{1}{N} \sum_{i=1}^{N/T} \E_{\zv_i }     \log  (P_z ||\zv_i||^2  +   \sigma^2  )      \nonumber \\
&=  \bigl(1-\frac{  1}{T} \bigr) \log(  1+ \frac{P_x}{ \sigma^2}) \!+\!  \frac{1}{N} \sum_{i=1}^{N/T} \E_{ \zv_i }  \log( 1+  \frac{P_x }{ \sigma^2 +  P_z ||\zv_i||^2}   ) .   \label{eq:1934}
\end{align}
This is the upper bound with the best Gaussian distributed signal $x$ and any distribution of the interference.

The above upper bound holds for any distribution of interference, and we now consider the worst case of the interference distribution $f_z$ for the upper bound in \eqref{eq:1934}. In this upper bound, only the second term $ \frac{1}{N}  \sum_{i=1}^{N/T} \E_{ \zv_i }  \log( 1+  \frac{P_x }{ \sigma^2 +  P_z ||\zv_i||^2}   )$ depends on $\zv$.  The function $f(t)=log(1+\frac{P_x}{\sigma^2+t P_z})$ is a convex function since $\frac{\partial^2 f(t)}{\partial t^2} \geq 0$. Therefore, we have
\begin{align}
 & \sum_{i=1}^{N/T} \E_{ \zv_i }  \log( 1+  \frac{P_x }{ \sigma^2 +  P_z ||\zv_i||^2}   )\\
&= \E_{\zv_1,\zv_2,\dots,\zv_{N/T}} \sum_{i=1}^{N/T}  \log( 1+  \frac{P_x }{ \sigma^2 +  P_z ||\zv_i||^2}   ) \\ \label{eq:jason_inequality}
&\geq \E_{\zv_1,\zv_2,\dots,\zv_{N/T}} \frac{N}{T}  \log( 1+  \frac{P_x }{ \sigma^2 +  P_z \frac{T}{N}\sum_{i=1}^{N/T}||\zv_i||^2}   ) \\  \label{eq:after_jason_inequality}
&=   \E_{\zv}  \left[ \frac{N}{T} \log( 1+  \frac{P_x }{ \sigma^2 +  \frac{T}{N} P_z  ||\zv||^2}   ) \right] \\
&=   \frac{N}{T} \log( 1+  \frac{P_x }{ \sigma^2 +  T P_z }   ) \label{eq:worst_interference}
\end{align}
%\begin{align}
% & \sum_{i=1}^{N/T} \E_{ \zv_i }  \log( 1+  \frac{P_x }{ \sigma^2 +  P_z ||\zv_i||^2}   )\\
%&= \sum_{i=1}^{N/T} \sum_{\zv_i} P_r(\zv_i)  \log( 1+  \frac{P_x }{ \sigma^2 +  P_z ||\zv_i||^2}   ) \\
%&= \sum_{i=1}^{N/T} \sum_{\zv_i} \sum_{\zv^N:\ \zv_i \; in \; \zv^N} P_r(\zv^N)  \log( 1+  \frac{P_x }{ \sigma^2 +  P_z ||\zv_i||^2}   ) \\
%&= \sum_{\zv^N} P_r(\zv^N) \sum_{i=1}^{N/T} \sum_{\zv_i:\ \zv_i \; in \; \zv^N}   \log( 1+  \frac{P_x }{ \sigma^2 +  P_z ||\zv_i||^2}   ) \\
%&= \sum_{\zv^N} P_r(\zv^N)  \sum_{ \begin{subarray} {c}
%i=1:N/t \\
%\zv_i:\ \zv_i \; in \; \zv^N
%\end{subarray}}
%\log( 1+  \frac{P_x }{ \sigma^2 +  P_z ||\zv_i||^2}   ) \\
%%&=\frac{1}{T}  \E_{ \zv^N } \left[ \sum_{i=1}^{N/T} \log( 1+  \frac{P_x }{ \sigma^2 +  P_z ||\zv_i||^2}   ) \right]\\
%&\geq   \sum_{\zv^N} P_r(\zv^N) \left[ \frac{N}{T} \log( 1+  \frac{P_x }{ \sigma^2 +  \frac{T}{N}P_z \sum_{ \begin{subarray} {c}
%i=1:N/t \\
%\zv_i:\ \zv_i \; in \; \zv^N
%\end{subarray}}
% ||\zv_i||^2}   ) \right] \\
%&=   \sum_{\zv^N} P_r(\zv^N) \left[ \frac{N}{T} \log( 1+  \frac{P_x }{ \sigma^2 +  \frac{T}{N} P_z  ||\zv^N||^2}   ) \right] \\ \label{eq:worst_interference}
%&=   \sum_{\zv^N} P_r(\zv^N) \left[ \frac{N}{T} \log( 1+  \frac{P_x }{ \sigma^2 +  T P_z }   ) \right]
%\end{align}
where eq. \eqref{eq:jason_inequality} follows the Jason inequality and the equality holds when $||\zv_i||^2=||\zv_j||^2$ for any $i,j$ that $1\leq i,j \leq N/T$, and eq. \eqref{eq:after_jason_inequality} holds when $N$ is large enough such that the distribution of $\zv$ in one packet is ergodic as $||\zv||^2= N\E |z[k]|^2=N$.  Substitute \eqref{eq:worst_interference} into \eqref{eq:1934}. Then, we obtain the final upper bound as
$$
\bigl(1-\frac{  1}{T} \bigr) \log(  1+ \frac{P_x}{ \sigma^2}) \!+\!  \frac{1}{T}  \log( 1+  \frac{P_x }{ \sigma^2 +  T P_z }   )
$$
where the signal has Gaussian distribution and the interference has constant power for each block.

\subsection{Link Between BKIC-ZF and Traditional KI Cancellation} \label{proof_link}

Traditionally, known interference is cancelled as follows. With reference to \eqref{eq:system_model_simple} and the constant power modulation model therein, the interference channel coefficient is first estimated by
\beq
\hat{h} = \frac{1}{T}\sum_{k=1}^T{r'[k]}.
\eeq
Then, the interference can be subtracted from the received signal as:
\beq
\begin{split}\label{eq:traditional_kic}
x'[k]&=r'[k]-\hat{h}=x'[k]+h-\hat{h}+n'[k]\\
&=x'[k] - \frac{1}{T}\sum_{i=1}^T{x'[i]+n'[i]}+n'[k]
%&=\frac{T-1}{T} x_1[k] - \frac{1}{T}\sum_{i\neq k}{x_1[i]+n_1[i]} + \frac{T-1}{T}n_1[k]
\end{split}
\eeq
For BKIC in \eqref{eq:cancellation_matrix}, an intuitive recovery scheme is to multiply the inverse matrix $Q^{-1}$ to both sides of \eqref{eq:cancellation_matrix}. However, the matrix $Q$ is not full rank and $Q^{-1}$  does not exist. To make it invertible, we append an artificial row vector   $a_T=[1 \quad 1 \quad \dots \quad 1]/T$ to $Q$. Then, we obtain a new matrix denoted as $Q_1=\left[ \begin{array}{c} Q  \\ a_T \end{array} \right]$. According to \eqref{eq:cancellation_matrix} and the definition of $Q_1$, we can obtain
\beq \label{eq:mimo_zf}
\begin{split}
Q_1(\xv'+\nv')&=\left[ \begin{array}{c} Q  \\ a_T \end{array} \right](\xv'+\nv')\\
&=\left[ \begin{array}{c} \yv'  \\ \frac{1}{T}\sum_{k=1}^T{x'[k]+n'[k]} \end{array} \right]\\
&= \left[ \begin{array}{c} \yv'  \\ 0 \end{array} \right]+\mathbf{v}
\end{split}
\eeq
where $\mathbf{v}$ is a column vector whose transpose is $[0  \dots 0 \frac{1}{T}\sum_{k=1}^T{x'[k]+n'[k]}$. Therefore, the artificial vector $a_T$ is equivalentl to an averaging process over the received block. When the block length $T$ is large, the only non-zero element in $\mathbf{v}$ is random variable with zero mean and small variance.

Now, we can rewrite \eqref{eq:mimo_zf} into a standard MIMO form as
\beq \label{eq:mimo_zf1}
\begin{split}
\left[ \begin{array}{c} \yv'  \\ 0 \end{array}  \right] = Q_1(\xv'+\nv')-\mathbf{v}
\end{split}
\eeq
with zero forcing MIMO detection scheme,    we can obtain the estimate of the target signal as
\beq
\widehat{\xv'+\nv'} = Q_1^{-1} \left[ \begin{array}{c} \yv'  \\ 0 \end{array}  \right]=\xv'+\nv' - Q_1^{-1}\mathbf{v}
\eeq
where the inverse of $Q_1$ exists and it is
$$
Q_1^{-1}=\frac{1}{T}\begin{bmatrix}
T-1 & T-2 & T-3 & \cdots & 1 & T \\
-1 & T-2 & T-3 & \cdots & 1 & T \\
-1 & -2 & T-3 & \cdots & 1 & T\\
-1 & -2 & -3 & \cdots &1 & T\\
\vdots &\cdots & \ddots & \ddots & \vdots \\

-1 & -2 & -3 & \cdots & -(T-1) & T
\end{bmatrix}$$
and the post detection noise is $- Q_1^{-1}\mathbf{v}$ and its transpose is $-[\frac{1}{T}\sum_{k=1}^T{x'[k]+n'[k]} \cdots \frac{1}{T}\sum_{k=1}^T{x'[k]+n'[k]} ]$. We refer this scheme as BKIC-ZF.

By comparing the formulation of the residual interference with the traditional KIC in \eqref{eq:traditional_kic}  and that in BKIC-ZF \eqref{eq:mimo_zf1}, we can obtain the following conclusion: The traditional KIC with Least Square channel estimator is exactly equivalent to the BKIC-ZF \cite{BKIC_JSAC:13}.

\subsection{Achievable Rate of Traditional KIC} \label{proof_tradtional}
We first review the traditional interference cancellation scheme and then calculate the achievable rate by assuming arbitrary distribution of the interference $z[k]$. With reference to \eqref{eq:system_model}, the interference channel coefficient is first estimated by
\beq
\begin{split}
\hat{h}& = \frac{1}{\sqrt{P_z}\sum_{i=1}^T{|z[i]|^2}}\sum_{i=1}^T{z^*[i]r[i]}\\
&=h+\frac{\sum_{i=1}^T{z^*[i](\sqrt{P_x}x[i]+n[i])}} {\sqrt{P_z}\sum_{i=1}^T{|z[i]|^2}}
\end{split}
\eeq
where the superscript $*$ denotes the conjugate operation for a scalar variable. Then, the interference part can be subtracted from each received symbol as follows:
\beq
\begin{split}\label{eq:appendix_cancellation}
&\sqrt{P_x}\hat{x}[k] = r[k]-\sqrt{P_z}\hat{h}z[k] \\
&=\sqrt{P_x}x[k]+\sqrt{P_z}(h-\hat{h})z[k]+n[k]\\
&=\sqrt{P_x}x[k]-\frac{z[k]}{\sum_{i=1}^T{|z[i]|^2}} \sum_{i=1}^T {z^*[i]\left[ \sqrt{P_x}x[i]+n[i]\right]}
%\hat{h}& = \frac{1}{\sqrt(P_z)\sum_{i=1}^T{|z[i]|^2}}\sum_{i=1}^T{z*[i]r[i]}\\
%&=h+\frac{\sum_{i=1}^T{z*[i](x[i]+n[i])}} {\sqrt(P_z)\sum_{i=1}^T{|z[i]|^2}}
\end{split}
\eeq
We can rewrite the above signal as
\beq
\begin{split}
&\frac{\sum_{i=1}^T{|z[i]|^2}}{\sum_{i\neq k}{|z[i]|^2}}   \sqrt{P_x}\hat{x}[k]  \\
&=\sqrt{P_x}x[k]+n[k]-\frac{z[k]}{\sum_{i\neq k}{|z[i]|^2}} \sum_{i\neq k} {z^*[i]\left[ \sqrt{P_x}x[i]+n[i]\right]}
%\hat{h}& = \frac{1}{\sqrt(P_z)\sum_{i=1}^T{|z[i]|^2}}\sum_{i=1}^T{z*[i]r[i]}\\
%&=h+\frac{\sum_{i=1}^T{z*[i](x[i]+n[i])}} {\sqrt(P_z)\sum_{i=1}^T{|z[i]|^2}}
\end{split}
\eeq
where the first term is the target signal and the last two terms are regarded as noise.  Then, the SINR of the $k$-th signal is
\beq \label{eq:app_sinr}
SINR_k=\frac{\sum_{i\neq k}{|z[i]|^2}P_x} {\sigma^2\sum_{i=1}^T{|z[i]|^2} + P_x|z[k]|^2}.
\eeq
Assuming Gaussian distribution for $x[k]$, the mutual information of this symbol is given by
\beq
\begin{split}
I[k]&=\log(1+SINR_k) \\
&= \log\left[\frac{(P_x+\sigma^2)\sum_{i=1}^T{|z[i]|^2}} {\sigma^2\sum_{i=1}^T{|z[i]|^2} + P_x|z[k]|^2} \right]\\
&=\log\left[\frac{(P_x+\sigma^2)} {\sigma^2 + P_x|z[k]|^2/A} \right]
\end{split}
\eeq
where $A=\sum_{i=1}^T{|z[i]|^2}$ is the normalized total power of the interference within one block.

We now obtain the smallest total mutual information, related to the worst distribution of  $z[k]$, in the same manner as in Appendix A.
\beq
\begin{split}
&I[k]+I[j]\\
&=\log\left[\frac{(P_x+\sigma^2)} {\sigma^2 + P_x|z[k]|^2/A}\right]  +  \log\left[\frac{(P_x+\sigma^2)} {\sigma^2 + P_x|z[j]|^2/A} \right]\\
&\leq 2\log\left[\frac{(P_x+\sigma^2)} {\sigma^2 + P_x(|z[k]|^2+|z[j]|^2)/2A} \right]
\end{split}
\eeq
Then, with the total interference power constraint, the total mutual information is minimized when each interfering symbol has the same power, i.e., $|z[k]|=1$ for all $k$. Finally, the achievable rate with traditional known interference cancellation scheme is
\beq
\begin{split}
R_t&=I=\log(\frac{P_x+\sigma^2}{P_x/T+\sigma^2})\\
&=\log(1+\frac{(T-1)\gamma}{T+\gamma}).
\end{split}
\eeq

\subsection{Achievable Rate with Orthogonal Training Sequence} \label{proof_orthogonal}
In the traditional known interference cancellation scheme, the performance is degraded due to the limited channel estimation accuracy. In order to improve the channel estimation accuracy, orthogonal training sequence is often used. Without loss of generality, we assume time orthogonality. Specifically, the transmitter is assumed to know the duration of the interference and set $x[T]=0$ so as to not affect the $T$-th interfering symbol in each block, without loss of generality. At the receiver, the interference channel of each block is first estimated with the $T$-th symbol as
\beq
\hat{h}=\frac{r[T]}{\sqrt{P_z}z[T]}=h+\frac{n[T]}{\sqrt{P_z}z[T]}.
\eeq
Then, this estimated channel coefficient is used to cancel the interference of the other symbols in the same block as in \eqref{eq:appendix_cancellation}. Since there are only $T-1$ information bearing symbols in one block, the achievable rate of this scheme is
\beq
R = (1-1/T)\log(1+\frac{P_x}{\sigma^2+\delta}).
\eeq
where $\delta$ is a small value depending on the channel estimation error $\frac{n[T]}{\sqrt{P_z}z[T]}$ . Compare to \eqref{eq:bkic_rate1}, we can find that our BKIC scheme can achieve a strictly better performance than this orthogonal training scheme with coordination.

%%   05/21/2014 -- JC
%\section{Appendix - Converse}

%\beq %\label{eq:system_model}
%\begin{split}
%r[k]&=\sqrt{P_x}x[k]+\sqrt{P_z}h[k] z[k]+n[k]
%\end{split}
%\eeq

%======================================
\bibliographystyle{IEEEtran}
\bibliography{IEEEabrv,final_refs}
%\bibliography{IEEEabrv,Known_interference_0708-jsac}

\end{document}